\documentclass[sn-nature,oneside,referee]{sn-jnl}

\usepackage{graphicx}%
\usepackage{multirow}%
\usepackage{amsmath,amssymb,amsfonts}%
\usepackage{amsthm}%
\usepackage{mathrsfs}%
\usepackage[title]{appendix}%
\usepackage{xcolor}%
\usepackage{textcomp}%
\usepackage{manyfoot}%
\usepackage{booktabs}%
\usepackage{xurl} 
\usepackage{algorithm}%
\usepackage{algorithmicx}%
\usepackage{algpseudocode}%
\usepackage{listings}%
\usepackage{chemformula}
\usepackage{siunitx}
\usepackage[utf8]{inputenc}
\usepackage{xr}

\unnumbered



\begin{document}

\title{Enhanced Charge-Density-Wave Order and Suppressed Superconductivity in Intercalated Bulk $\mathrm{Nb}{\mathrm{Se}}_{2}$}

\author*[1]{\fnm{Huanhuan} \sur{Shi}\email{huanhuan.shi@kit.edu}}
\author[2]{\fnm{Qili} \sur{Li}}
\author[1,3]{\fnm{Antoine} \sur{M. T. Baron}}
\author[3]{\fnm{Marie-Aude} \sur{Méasson}}
\author[4,5,6]{\fnm{Sangjun} \sur{Kang}}
\author[1]{\fnm{Dirk} \sur{Fuchs}}
\author[1]{\fnm{Fabian} \sur{Henssler}}
\author[2]{\fnm{Alexander} \sur{Haas}}
\author[1]{\fnm{Paolo} \sur{Battistoni}}
\author[1]{\fnm{Nour} \sur{Maraytta}}
\author[1,6]{\fnm{Michael} \sur{Merz}}
\author[1]{\fnm{Amir-Abbas} \sur{Haghighirad}}
\author[1,2]{\fnm{Wulf} \sur{Wulfhekel}}
\author[4,5,6]{\fnm{Christian} \sur{Kübel}}
\author*[1]{\fnm{Matthieu} \sur{Le Tacon}}\email{matthieu.letacon@kit.edu}

\affil[1]{\orgdiv{Institute for Quantum Materials and Technologies}, 
  \orgname{Karlsruhe Institute of Technology}, 
  \orgaddress{\street{Kaiserstr.\ 12}, \postcode{76131}, \city{Karlsruhe}, \country{Germany}}}

\affil[2]{\orgdiv{Physikalisches Institut}, 
  \orgname{Karlsruhe Institute of Technology}, 
  \orgaddress{\street{Wolfgang-Gaede-Str.\ 1}, \postcode{76131}, \city{Karlsruhe}, \country{Germany}}}

\affil[3]{\orgname{Institut Néel CNRS/UGA UPR2940}, 
  \orgaddress{\street{Rue des Martyrs 25}, \postcode{38042}, \city{Grenoble}, \country{France}}}

\affil[4]{\orgdiv{Institute of Nanotechnology}, 
  \orgname{Karlsruhe Institute of Technology}, 
  \orgaddress{\street{Kaiserstr.\ 12}, \postcode{76131}, \city{Karlsruhe}, \country{Germany}}}

\affil[5]{\orgdiv{In-situ Electron Microscopy, Department of Materials and Earth Sciences}, 
  \orgname{Technische Universität Darmstadt}, 
  \orgaddress{\street{Karolinenpl.\ 5}, \postcode{64289}, \city{Darmstadt}, \country{Germany}}}

\affil[6]{Karlsruhe Nano Micro Facility (KNMFi), Karlsruhe Institute of Technology, Kaiserstr. 12, 76131 Karlsruhe, Germany}

\date{\today}

\abstract{The electronic ground states of transition-metal dichalcogenides are strongly shaped by reduced dimensionality, yet the properties of atomically thin layers remain difficult to probe due to their small size and environmental sensitivity. Here we demonstrate that controlled electrochemical intercalation of organic cations provides a robust bulk platform for accessing monolayer-like physics in NbSe$_2$. Intercalation of tetrapropylammonium and tetrabutylammonium expands the interlayer spacing by nearly a factor of two, electronically decoupling the NbSe$_2$ layers while simultaneously introducing well-defined charge doping. Using a combination of Raman spectroscopy, scanning tunneling microscopy, X-ray diffraction, and photoemission, we uncover a pronounced enhancement of the charge–density–wave transition temperature to $\sim \SI{130}{K}$ together with a strong suppression of superconductivity, reproducing the phase diagram observed in exfoliated monolayers. The enhanced  charge–density–wave order and reduced $T_c$ arise from the combined effects of dimensionality reduction and electron injection, and are accompanied by distinct dip–hump anomalies in the tunneling spectra suggestive of collective mode excitations. Our results establish molecular intercalation as a powerful and scalable route for engineering competing orders in layered quantum materials.}

\maketitle

\section*{Introduction}

The reduced dimensionality of monolayer transition-metal dichalcogenides (TMDs) gives rise to distinctive physical properties absent in bulk crystals, spurring intense research activity in the field of low-dimensional materials~\cite{Manzeli_2017,Wang_2012}. At the monolayer limit, 
direct band gaps in the visible range, strong spin–orbit coupling, and pronounced excitonic effects emerge, enabling phenomena and applications that are inaccessible in the bulk compound.

Bulk TMD 2H-NbSe$_2$ possesses a hexagonal $P6_3/mmc$ structure and hosts both a charge–density–wave (CDW) order and superconductivity, emerging below 
$T_{\mathrm{CDW}} = 33$\,K and $T_c = 7.2$\,K, respectively. The interplay between these two orders has been extensively investigated~\cite{Leroux_PRB2015,Sanna_2022,Cho_NC_2018,Majumdar_PRM_2020}. Beyond its rich phase diagram, 2H-NbSe$_2$ serves as a model system to study CDW formation in dimensions $d > 1$, where the canonical Peierls picture of Fermi-surface-nesting–driven CDWs breaks down. Instead, the nearly commensurate instability at $q_{\mathrm{CDW}} = (0.329, 0, 0)$~\cite{Moncton_1975}, responsible for the characteristic $\sim 3 \times 3$ superstructure, was shown to originate from anisotropic electron–phonon coupling~\cite{Weber_PRL2011,Johannes_2006}. The superconductivity in this material is also believed to be conventional in nature~\cite{Valla_PRL_2004}, arising from the same electron–phonon interaction~\cite{Leroux_PRB2015}, making NbSe$_2$ a valuable platform for exploring the microscopic coupling between CDW and superconducting states.

A particularly appealing aspect of NbSe$_2$ is its tunability. External pressure~\cite{Berthier_SSC1976,Suderow_2005,Feng_PNAS2012,Leroux_PRB2015}, 
strain~\cite{Gao_2018}, structural modifications such as misfit phases~\cite{Rouxel_1995,Leriche_AFM2021,Zullo_2024}, and, most prominently, reduction in dimensionality to the monolayer limit~\cite{Ugeda_2016,Xi_2015}, all strongly reshape the balance between CDW and superconductivity, offering important insights into their microscopic origin.

The monolayer NbSe$_2$ prepared by mechanical exfoliation provides an ideal platform for examining these competing orders in the two-dimensional limit~\cite{Prakiran_2021,Cao_2015}. Owing to its non-centrosymmetric structure and strong spin–orbit coupling, monolayer NbSe$_2$ hosts Ising superconductivity, a phenomenon absent in the bulk~\cite{Xi_2016}. In this regime, $T_c$ is strongly reduced to $0.9$–$3.7$\,K, while $T_{\mathrm{CDW}}$ increases rapidly, reaching $145$ \,K~\cite{Xi_2015}. In particular, this enhancement occurs without a change in the CDW wave vector~\cite{Ugeda_2016,Nakata_2018}. The opposing evolution of $T_c$ and $T_{\mathrm{CDW}}$ reinforces the competitive relationship between CDW formation and superconductivity. Nevertheless, the microscopic mechanisms driving CDW order and superconductivity~\cite{zhang_PhysRevB_2019,Wickramaratne_PhysRevX_2020}, and the 
origins of their interplay~\cite{Johannes_2006,Feng_PNAS2012,Leroux_PRB2015,Moulding_PRR2020}, 
remain the subject of active debate.

Despite their utility, exfoliated monolayers suffer from intrinsic limitations, including micron-scale lateral dimensions, time-consuming preparation, and air sensitivity, which restrict the range of experimental probes that can be reliably applied. To circumvent these challenges, chemical intercalation has emerged as an effective strategy to suppress interlayer coupling in bulk crystals and induce monolayer-like behavior without the need for exfoliation~\cite{Wan_2024}. 
Beyond decoupling layers, intercalated ions can donate charge carriers, providing simultaneous control over dimensionality and electron density~\cite{Dresselhaus_1981}.

This approach has recently been brought to bear on NbSe$_2$. For example, Zhang \textit{et al.} demonstrated that ionic-liquid intercalation results in a suppressed $T_{\mathrm{CDW}} \approx 68$\,K and enhanced $T_c \approx 6.9$\,K compared to monolayer samples~\cite{Zhang_2022}. Similarly, tetraheptylammonium-intercalated NbSe$_2$ exhibits $T_{\mathrm{CDW}} = 65$\,K and $T_c = 7.1$\,K~\cite{Yu_2024}. These observations further support the competitive nature of CDW and superconductivity. However, intercalated NbSe$_2$ has not yet displayed the concurrent enhancement of $T_{\mathrm{CDW}}$ and suppression of $T_c$ that is characteristic of the true monolayer limit.

\begin{figure}[h]
    \centering
    \includegraphics[width=0.68\linewidth]{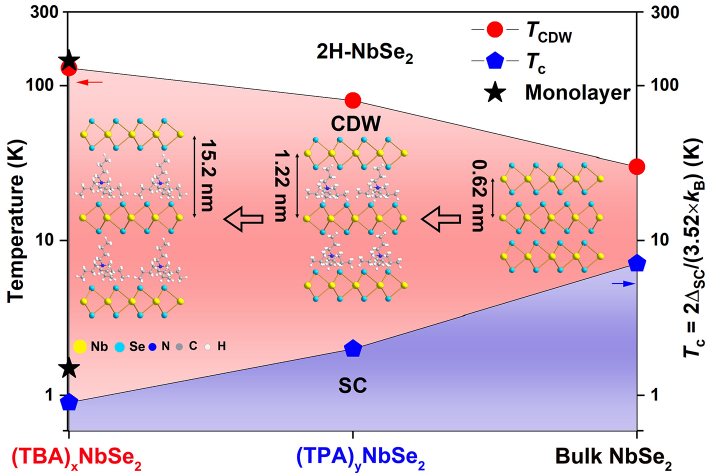}
    \caption{Intercalant–temperature phase diagram of 2H-NbSe$_2$ based on Raman  spectroscopy ($T_{\mathrm{CDW}}$) and STM (from $\Delta_{\text{SC}}$, see text); monolayer values are taken from Ref.~\cite{Xi_2015}. }
    \label{fig1}
\end{figure}

In this work, we synthesize large-area ($\sim$1\,mm) intercalated NbSe$_2$ crystals using variable-size organic cations (tetrapropylammonium (TPA$^+$) and tetrabutylammonium (TBA$^+$)) to systematically tune the interlayer spacing and carrier concentration. X-ray diffraction (XRD) and high-angle annular dark-field scanning transmission electron microscopy (HAADF-STEM) confirm that intercalation expands the interlayer spacing from \SI{0.62}{\nano\metre} in the bulk to \SI{1.22}{\nano\metre} in (TPA)$_y$NbSe$_2$ and \SI{1.52}{\nano\metre} in (TBA)$_x$NbSe$_2$, effectively decoupling adjacent layers. Raman spectroscopy reveals a pronounced enhancement of $T_{\mathrm{CDW}}$ to 70\,K in (TPA)$_y$NbSe$_2$ and a record-high 130\,K in (TBA)$_x$NbSe$_2$, while scanning tunneling microscopy (STM) shows a strong suppression of the superconducting gap ($\Delta_{\text{SC}}$), corresponding to reduced $T_c$ values of 2\,K and 0.9\,K, respectively (Fig.~\ref{fig1}). Notably, (TBA)$_x$NbSe$_2$ exhibits behaviors closely resembling those of exfoliated monolayers. These results further support the intrinsic competition between CDW formation and superconductivity in NbSe$_2$ and establish molecular intercalation as a powerful route to access monolayer-like physics in bulk crystals.

\section*{Results}
The structural effects of molecular intercalation (detailed in the Methods section) were investigated using a combination of Raman spectroscopy, XRD, and TEM. Reference Raman spectra obtained from pristine 2H-$\mathrm{NbSe}_{2}$ single crystals are presented in Supplementary Information (Fig.~S6) and exhibit all characteristic phonon modes previously reported at room temperature for this material. These include the A$_{1g}$ mode (out-of-plane vibrations) at \SI{234}{\per\centi\metre}, the E$^1_{2g}$ mode (in-plane vibrations) at \SI{250}{\per\centi\metre}, the soft mode (associated with a second-order scattering process) at \SI{181}{\per\centi\metre}, and the E$^2_{2g}$ shear mode at \SI{30}{\per\centi\metre}. These features are well established in the literature for bulk $\mathrm{NbSe}_{2}$~\cite{Grasset_PhysRevB2018}.

\begin{figure}[h]
    \centering
    \includegraphics[width=\linewidth]{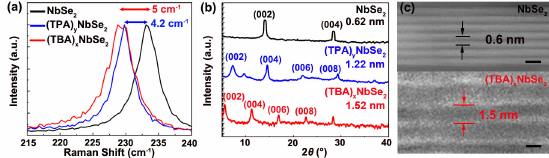}
    \caption{(a) A$_{1g}$ mode of NbSe$_2$, (TPA)$_y$NbSe$_2$ and (TBA)$_x$NbSe$_2$; (b) XRD patterns of  $\mathrm{Nb}{\mathrm{Se}}_{2}$, (TPA)$_y$NbSe$_2$ and (TBA)$_x$NbSe$_2$; (c) Cross-sectional HAADF-STEM images of pristine $\mathrm{Nb}{\mathrm{Se}}_{2}$ (top) and intercalated (TBA)$_x$NbSe$_2$ (bottom). Scale bars, \SI{1}{\nano\metre}. }
    \label{fig2}
\end{figure}

A direct signature of intercalation is shown in Fig.~\ref{fig2}-a, where a systematic redshift of the A$_{1g}$ mode is observed in the intercalated compounds (see full spectra in Supplementary Information Fig.~S7). Notably, the phonon lineshape remains similar to that of the pristine sample, indicating good structural integrity after intercalation. Relative to pristine 2H-$\mathrm{NbSe}_{2}$, the A$_{1g}$ mode softens by \SI{4.2}{\per\centi\metre} in (TPA)$_y$NbSe$_2$ and by \SI{5}{\per\centi\metre} in (TBA)$_x$NbSe$_2$, consistent with lattice expansion along the $c$-axis. The additional \SI{0.8}{\per\centi\metre} redshift observed in (TBA)$_x$NbSe$_2$ relative to (TPA)$_y$NbSe$_2$ suggests that the larger TBA$^+$ cation is more effective in weakening interlayer coupling in $\mathrm{NbSe}_{2}$.  

More importantly, a substantial hardening of \SI{9}{\per\centi\metre} in the E$^1_{2g}$ mode, as well as a complete suppression of the shear mode, is observed in the intercalated samples. These behaviors closely mirror those of the monolayer $\mathrm{NbSe}_{2}$~\cite{Xi_2015,Tan_2012}. Taken together, these results strongly suggest that the intercalated compounds effectively replicate the physical characteristics of the monolayer $\mathrm{NbSe}_{2}$.

Building on the Raman evidence for weakened interlayer coupling, we turn to XRD and cross-sectional TEM to quantitatively determine the interlayer separation and confirm the configuration adopted by the intercalated species. Indeed, tetraalkylammonium (TAA$^+$) ions can exist in two primary configurations: a flattened form and a tetrahedral form (see Supplementary Information Fig.~S8)~\cite{Roth_2021}. Fig.~\ref{fig2}-b compares the XRD patterns of the pristine and intercalated samples. The pristine crystal $\mathrm{NbSe}_{2}$ exhibits sharp diffraction peaks (00$\ell$), corresponding to an interlayer spacing of \SI{0.62}{\nano\metre}, in agreement with previous reports~\cite{Yu_2024}. Upon intercalation, the (00$\ell$) reflections of the pristine $\mathrm{NbSe}_{2}$ disappear entirely, and a new set of (00$\ell$) peaks appears at lower 2$\theta$ angles in the intercalated $\mathrm{NbSe}_{2}$, indicating successful intercalation of TAA$^+$ into the bulk crystal.
According to XRD results, the interlayer distances of (TPA)$_y$NbSe$_2$ and (TBA)$_x$NbSe$_2$ are \SI{1.22}{\nano\metre} and \SI{1.52}{\nano\metre} respectively. Compared to pristine $\mathrm{NbSe}_{2}$, the interlayer distances expand by approximately \SI{0.6}{\nano\metre} for (TPA)$_y$NbSe$_2$ and \SI{0.9}{\nano\metre} for (TBA)$_x$NbSe$_2$. Notably, the expansion of  (TPA)$_y$NbSe$_2$ is larger than a flattened form (0.4 nm) but smaller than a tetrahedral form (0.78 nm) of TPA$^+$, suggesting the presence of a distorted tetrahedral configuration of TPA$^+$. In contrast, the expanded interlayer spacing observed for (TBA)$_x$NbSe$_2$ is in good agreement with the tetrahedral TBA$^+$ cation, providing strong evidence that intercalation occurs by insertion of a single TBA$^+$ ion into its tetrahedral conformation (Fig.~\ref{fig1}). This conformation is further corroborated by cross-sectional HAADF-STEM imaging (Fig.~\ref{fig2}-c and Supplementary Information Fig. S9), which reveals an ordered lamellar structure with a periodicity in agreement with the XRD analysis. 

\begin{figure}[h]
    \centering
    \includegraphics[width=0.9\linewidth]{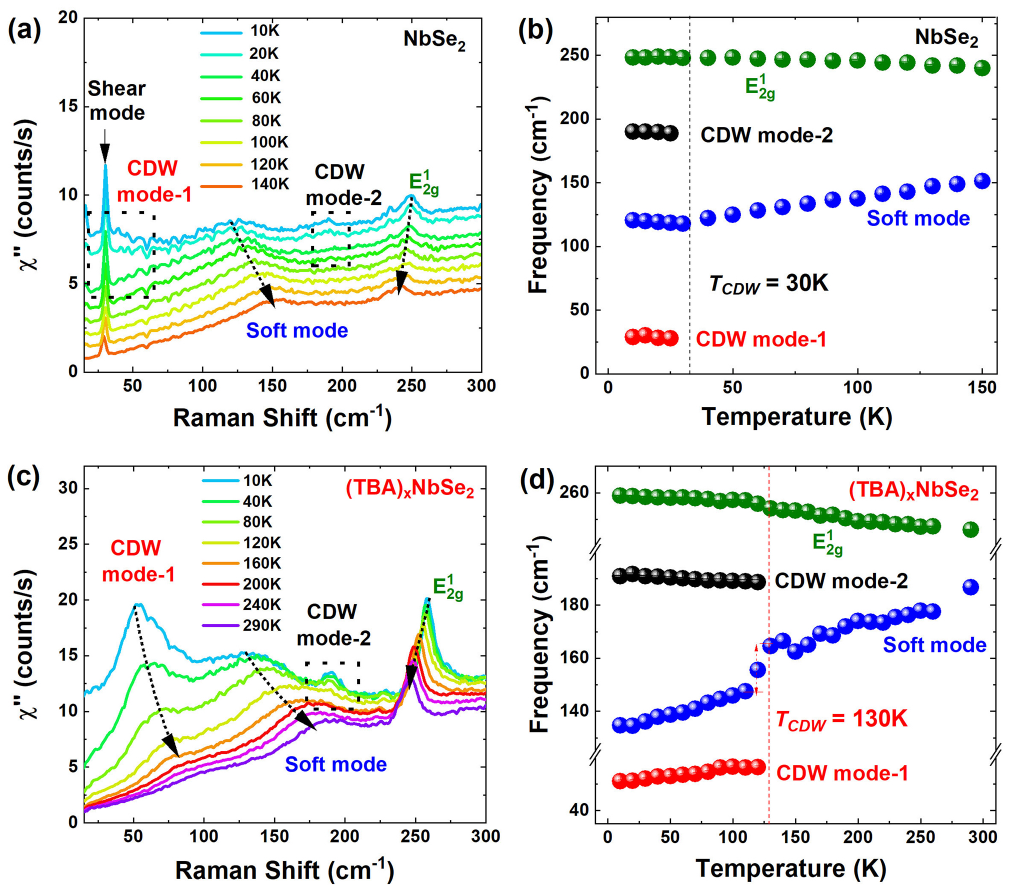}
    \caption{(a,c) Temperature-dependent Raman spectra of pristine NbSe$_2$ (a) and intercalated (TBA)$_x$NbSe$_2$ (c) in the ab polarization configuration; (b,d) Temperature dependence of the phonon frequencies in pristine NbSe$_2$ (b) and intercalated (TBA)$_x$NbSe$_2$ (d).}
    \label{fig3}
\end{figure}

Having determined both the conformation of the intercalated TAA$^+$ cations and the associated lattice expansion, we now turn to the effect of intercalation on the CDW properties of $\mathrm{NbSe}_{2}$. Fig~\ref{fig3} and Fig. S10 present temperature-dependent Raman spectra of pristine and intercalated $\mathrm{NbSe}_{2}$ measured between \SI{10}{\kelvin} and \SI{290}{\kelvin}. The $ab$ polarization configuration was used to minimize the background of elastic scattering, and all spectra were corrected for by the temperature-dependent Bose factor~\cite{Sen_2020}.

In pristine $\mathrm{NbSe}_{2}$, the soft mode redshifts upon cooling and saturates at \SI{117.8}{\per\centi\metre} below the CDW transition temperature $T_{\mathrm{CDW}}$ = \SI{33}{\kelvin} (Fig.~\ref{fig3}-a). At lower temperatures, two additional Raman modes emerge at \SI{28}{\per\centi\metre} (amplitude mode) and \SI{189}{\per\centi\metre}, signaling the onset of CDW order, consistent with previous reports~\cite{Tsang_1976,Sooryakumar_1980,Measson_PRB2014}. Strikingly, these CDW-associated modes appear at significantly higher temperatures in  intercalated $\mathrm{NbSe}_{2}$ (Fig.~\ref{fig3}-c and Supplementary Information Fig.~S10). The high-frequency mode remains near \SI{189}{\per\centi\metre}, while the amplitude mode blue-shifts to \SI{60}{\per\centi\metre} in (TPA)$_y\mathrm{NbSe}_{2}$ and \SI{56}{\per\centi\metre} in (TBA)$_x\mathrm{NbSe}_{2}$, reflecting enhanced electron–phonon coupling~\cite{Xi_2015}. In particular, the amplitude mode in (TBA)$_x\mathrm{NbSe}_{2}$ can survive to higher temperatures than in (TPA)$_y\mathrm{NbSe}_{2}$.

To quantitatively extract $T_{\mathrm{CDW}}$, the spectra were fitted using a damped harmonic oscillator model (see Supplementary Information Fig.~S11-12)~\cite{Cao_2023}. As shown in Fig.~\ref{fig3}-b, the soft mode in pristine $\mathrm{NbSe}_{2}$ softens from 181 to \SI{118}{\per\centi\metre} upon cooling to $T_{\mathrm{CDW}}$, below which it stabilizes. A similar trend is observed in (TBA)$_x\mathrm{NbSe}_{2}$:
the soft mode frequency follows a bulk-like behavior to \SI{130}{\kelvin}, at which point it abruptly changes from \SI{164.4}{\per\centi\metre} to \SI{147.5}{\per\centi\metre}. This discontinuity is accompanied by anomalies in the A$_{1g}$ and E$^1_{2g}$ phonon modes (see Supplementary Information Figs.~S13-15). Concurrently, CDW modes emerge, gain intensity, harden and sharpen below \SI{130}{\kelvin}, marking the onset of CDW order. These results clearly identify $T_{\mathrm{CDW}} \approx 130$\,K in (TBA)$_x\mathrm{NbSe}_{2}$, showing the highest reported $T_{\mathrm{CDW}}$ in intercalated $\mathrm{NbSe}_{2}$ to date~\cite{Zhang_2022,Yu_2024}.

\begin{figure}[h]
    \centering
    \includegraphics[width=\linewidth]{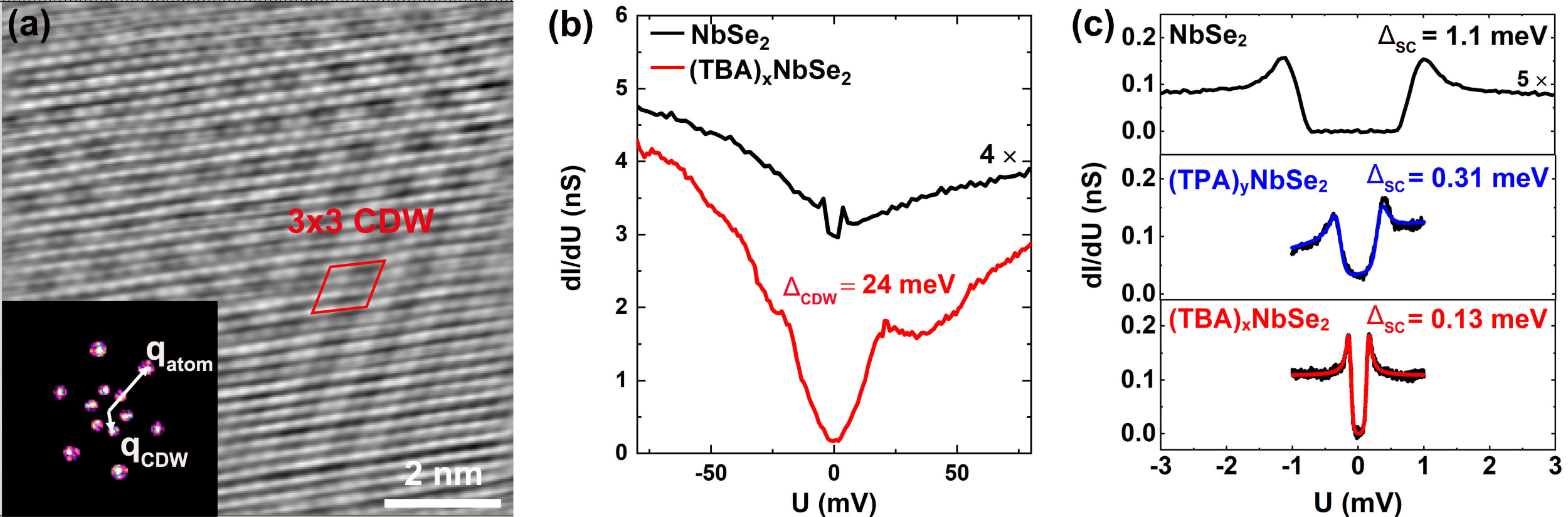}
    \caption{(a) The STM image of the intercalated (TBA)$_x$NbSe$_2$. The tip was stabled at sample bias 20 mV with tunneling curren 20 pA; (b-c) CDW gap and superconductivity gap of NbSe$_2$ and intercalated (TBA)$_x$NbSe$_2$ measured at 44 mK. Measuring conditions: (a) the tip was stabled at sample bias $\SI{20}{\mV}$ with tunneling current $\SI{20}{\pA}$; (b) the tip was stabled at sample bias $\SI{100}{\mV}$ and tunneling current $\SI{100}{\pA}$, and the lock-in amplifier had a modulation amplitude $\SI{4}{\mV}$; (c) the tip was stabled at sample bias $\SI{5}{\mV}$ and tunneling current $\SI{100}{\pA}$ with modulation amplitude $\SI{50}{\uV}$ for NbSe$_2$, and sample bias $\SI{1}{\mV}$ and tunneling curren $\SI{100}{\pA}$ with modulation amplitude $\SI{15}{\uV}$ for (TPA)$_y$NbSe$_2$ and (TBA)$_x$NbSe$_2$. }
    \label{fig4}
\end{figure}

To directly visualize the CDW order in real space, we performed STM measurements on (TBA)$_x\mathrm{NbSe}_{2}$. Fig.~\ref{fig4}-a displays representative STM topographs together with their two-dimensional Fourier transforms. A nearly perfect hexagonal atomic lattice is resolved, along with a well-defined, long-range incommensurate $(\sim 3 \times 3)$ CDW modulation.  In contrast to metal-intercalated NbSe$_2$, where defects and disorder are often introduced~\cite{Chatterjee_2015}, intercalation with TBA$^+$ produces no observable structural defects (see Supplementary Information Fig.~S17) and preserves the integrity of CDW. The unit cell of the superlattice closely resembles that of pristine $\mathrm{NbSe}_{2}$~\cite{Sanna_2022}, attesting to the minimal perturbation of the host lattice by the organic cation. Furthermore, compared to $\mathrm{NbSe}_{2}$, a wide CDW gap ($\Delta_{\text{CDW}}$) of about \SI{24}{\meV} emerges near the Fermi level in (TBA)$_x\mathrm{NbSe}_{2}$ (Fig.~\ref{fig4}-b). That may be caused by partial Fermi surface gapping, multiband effects, and strong coupling mechanisms~\cite{Ugeda_2016, Wang_1990}. 
Using the measured CDW gap ($\Delta_{\mathrm{CDW}} \approx 24~\mathrm{meV}$), we obtain
\[
\frac{2\Delta_{\mathrm{CDW}}}{k_{\mathrm{B}} T_{\mathrm{CDW}}} \approx 4.4,
\]
exceeding the canonical weak–coupling BCS/Peierls value of 3.52~\cite{Bardeen_1956}. This elevated ratio is a hallmark of strong electron–phonon coupling and is often associated with CDW states that deviate from a simple nesting–driven Peierls picture, consistent with prior reports on NbSe$_2$~\cite{Chatterjee_2015, Borisenko_2009}.

In addition, we investigated the effect of cation intercalation on superconductivity using both STM and Raman spectroscopy. The STM measurements clearly show a well-defined $\Delta_{\text{SC}}$ of approximately \SI{0.31}{meV} for (TPA)$_y\mathrm{NbSe}_{2}$ and \SI{0.13}{meV} for (TBA)$_x\mathrm{NbSe}_{2}$ (Fig.~\ref{fig4}-c), which is about an order of magnitude smaller than gap sizes (\SI{1.1}{meV}) in the pristine $\mathrm{NbSe}_{2}$~\cite{Sanna_2022}. Assuming the weak coupling limit of BCS, the superconducting $T_c$ 
is around $\SI{2}{\K}$ for (TPA)$_y\mathrm{NbSe}_{2}$ and $\SI{0.9}{\K}$ for (TBA)$_x\mathrm{NbSe}_{2}$, respectively, indicating a strong suppression induced by cation intercalation. This suppression is in agreement with Raman data measured down to $\SI{2}{\K}$ which shows no signature of superconductivity in (TBA)$_x\mathrm{NbSe}_{2}$. Indeed, the sharp superconducting mode that appears near \SI{19}{\per\centi\metre} in both $aa$ and $ab$ polarizations in pristine $\mathrm{NbSe}_{2}$~\cite{Sooryakumar_1980} is completely absent in (TBA)$_x\mathrm{NbSe}_{2}$ (see Supplementary Information Fig.~S18). 

\section*{Discussion}

The strong suppression of $T_c$ observed in intercalated compounds may result from significant changes in carrier concentration introduced during the electrochemical process. 
The nominal electron density is estimated to be $\sim 2.1 \times 10^{14}\,\mathrm{cm}^{-2}$  (see Supplementary Information for details), corresponding to about $10\%$ of the intrinsic carrier density in pristine NbSe$_2$~\cite{Xi_PRL2016}. This is consistent with several independent spectroscopic signatures, including the shift of the density-of-states peaks toward negative bias in STM measurements and the redshift of the Nb~$3d$ and Se~$3d$ core levels (see Supplementary Information Figs.~S19–S20), both of which indicate electron injection into the NbSe$_2$ layers.

Interestingly, larger electron doping achieved in NbSe$_2$ bilayers through misfit phases has recently been shown to suppress the CDW instability~\cite{Zullo_2024}. The contrast with our observations raises the question of whether the distinct behavior originates from the different doping amplitudes, or instead reflects qualitatively different responses in monolayers and bilayers. Addressing these possibilities will be an important direction for future study.

Further insight into the superconducting state is provided by the tunneling spectra, which display pronounced dip–hump anomalies at both positive and negative bias beyond the superconducting gap (see Supplementary Information Fig.~S21). These features appear at an energy scale compatible with interband Josephson coupling and closely resemble the signatures recently associated with Leggett collective modes in single-layer NbSe$_2$~\cite{Leggett_1966,Wan_2022}. While a definitive identification will require momentum-resolved probes and theoretical modeling, the similarity in phenomenology lends support to this interpretation. More broadly, these results underscore the potential of intercalated bulk crystals as a platform for exploring collective excitations in multiband superconductors.

\section*{Conclusion}

In summary, we have shown that electrochemical intercalation of TPA$^{+}$ and TBA$^{+}$ ions into bulk NbSe$_2$ produces a substantial expansion of the interlayer spacing and an effective electronic decoupling of the constituent layers. This structural modification leads to a pronounced enhancement of the CDW transition temperature, reaching $T_{\mathrm{CDW}} \simeq 130$~K, together with a strong suppression of superconductivity. The resulting phase diagram closely mirrors that of exfoliated monolayers~\cite{Zhang_2022,Nakata_2021}, demonstrating that molecular intercalation provides a robust and scalable route for accessing monolayer-like electronic ground states in macroscopic bulk crystals.

The concurrent enhancement of CDW order and reduction of $T_c$ highlight the intrinsic competition between these two collective phenomena in NbSe$_2$. Notably, the magnitude of the effects observed here exceeds that reported in mechanically exfoliated monolayers, suggesting that intercalation introduces additional tuning parameters—most prominently electron doping—that further reinforce CDW order. The quantitative determination of the carrier density, supported by XPS and STM measurements, confirms the important role played by charge transfer in steering the balance between competing ground states.

Finally, the observation of dip–hump anomalies consistent with Leggett-mode–like collective excitations reveals new opportunities to investigate multiband dynamics and superconducting collective modes in the two-dimensional limit. Overall, molecular intercalation establishes a powerful platform for engineering dimensionality, carrier concentration, and competing orders in layered quantum materials, offering a promising pathway toward unraveling the microscopic 
mechanisms that govern unconventional charge-ordered superconductors.

\section*{Methods and Materials}
\subsection{Synthesis of NbSe$_2$ single crystal}

NbSe$_2$ Single crystals were grown by chemical vapour transport (CVT) using iodine ($\SI{99.8}{\%}$, Riedel de Haën) as the transport agent. A mixture consisting of $\SI{2}{\g}$ prereacted NbSe$_2$, $\SI{120}{\mg}$ of iodine and an additional $\SI{11.5}{\mg}$ of selenium ($\SI{99.9}{\%}$, Roth) was sealed under vacuum in a fused silicon ampoule. The ampoule was then placed in a two-zone furnace, with the source and sink temperatures maintained at $\SI{810}{^\circ C}$ and $\SI{750}{^\circ C}$, respectively, to facilitate directed transport. After 12 days of growth under these conditions, large plate-shaped NbSe$_2$ crystals were obtained at the cooler end of the ampoule. Single crystals were achieved by washing with ethanol to remove iodine remains (see Supplementary Information Figs.~S2-3). 
 
\subsection{Intercalation of  (TAA)$_x$ molecules }
The intercalated $\mathrm{Nb}{\mathrm{Se}}_{2}$ crystals were synthesized via an electrochemical intercalation process using a two-electrode system~\cite{Shi_2023,Shi_2020} (see Supplementary Information Figs.~S4), where the $\mathrm{Nb}{\mathrm{Se}}_{2}$ crystal acted as cathode and a platinum foil was used as anode. The cathode and anode were placed in parallel with a constant distance of $\SI{1.0}{\cm}$. The electrolyte was made of tetraalkylammonium bromide (referred to hereafter as TAA$^+$, alkyl = propyl, butyl) and anhydrous dimethylformamide (DMF, $0.01\,\mathrm{M}$, $\SI{5}{\ml}$). With a low current ($20$-$80\,\mu\text{A}$), organic cations such as TPA$^+$ and TBA$^+$ are capable of intercalating in van der Waals gaps of the $\mathrm{Nb}{\mathrm{Se}}_{2}$ crystal (see Supplementary Information Fig.~S4). By varying in size, the intercalated cations enable control over interlayer spacing and doping levels, making them a valuable tool to tune the electronic properties of $\mathrm{Nb}{\mathrm{Se}}_{2}$. The associated electrochemical reactions are outlined in the following equations:

\begin{align*}
\text{Anode:} & \quad x\mathrm{Br}^- - x e^- \rightarrow \frac{x}{2} \mathrm{Br}_2 \\
\text{Cathode:} & \quad \mathrm{NbSe}_2 + x e^- + x\mathrm{TAA}^+ \rightarrow (\mathrm{TAA})_x\mathrm{NbSe}_2
\end{align*}

The compositions of the intercalated crystals were determined to be (TPA)$_{0.13}$NbSe$_2$ and (TBA)$_{0.22}$NbSe$_2$ based on the number of transferred electrons and energy-dispersive X-ray spectroscopy
(see Supplementary Information Section I, Table S1 and Figs. S1, S5 and S16). We could further check that the intercalated $\mathrm{NbSe}_{2}$ samples remain stable for at least two months (see Supplementary Information Fig.~S22), in contrast to the monolayer counterpart, which degrades within only a few hours~\cite{wang_NC_2023}.

\subsection{Low-temperature Raman spectroscopy}
Polarized Raman scattering was performed using a Horiba Jobin-Yvon LabRAM HR Evolution spectrometer equipped with a liquid-nitrogen-cooled CCD detector for measurements from 10 to $\SI{290}{\K}$. A He–Ne laser ($\lambda = 632.8$\,nm) with power kept below 1\,mW was focused on the sample through a 50$\times$ objective, producing a spot size of $\sim$5\,$\mu$m. To suppress elastically scattered light, the detection path included one notch filter and two Bragg filters. Spectra were acquired using \SI{1800}{\per\milli\metre} and \SI{600}{\per\milli\metre} gratings, corresponding to spectral resolutions of \SI{0.6}{\per\centi\metre} and \SI{1.6}{\per\centi\metre}, respectively. Superconductivity mode was investigated using Trivista triple-grating Raman spectrometer with an incident laser line at \SI{532}{\nano\metre}. Measurements were taken at $\SI{2}{\K}$ with a $^4$He closed-cycle cryostat and and with low power ($\SI{0.05}{\mW}$) to reduce laser heating effect. All spectra were corrected from the Bose thermal population factor.  

To probe excitations of different symmetries, we employed two polarization configurations: parallel ($aa$) and perpendicular ($ab$), in which the incident light polarization lies along the crystallographic $a$ axis and is either parallel or perpendicular to the polarization of the scattered light, respectively. According to Raman selection rules~\cite{Hayes1978}, these geometries allow access to the A$_{1g}$+E$_{2g}$ and E$_{2g}$ symmetry channels, respectively, in the hexagonal phase (point group D$_{6h}^1$).

\subsection{Scanning tunneling microscopy}
STM measurements were performed in a home-built ultra-high vacuum chamber housing a slider-type STM at milli-Kelvin temperatures \cite{Balashov_2018}. The base pressure during the experiments was \SI{1e-10}{\milli\bar}. The intercalated NbSe$_2$ was cleaved under ultrahigh vacuum with Kapton tape at room temperature and then immediately transferred to STM. All STM images were recorded in constant current mode. The W tip was initially prepared by flashing in the preparation chamber and was controlled in STM by dipping it into an Au(111) substrate, as well as by voltage pulsing on NbSe$_2$. A lock-in amplifier was used for spectroscopy.


\newpage

\clearpage

\backmatter
\bmhead{Acknowledgements}
H. Shi thanks S. Souliou, M. Frachet, and C. Singh for fruitful discussion and is grateful to the Karlsruhe Nano-Micro Facility (KNMFi) for technical support. We acknowledge the funding by the European Research Council (ERC) under the European Union's Horizon 2020 research and innovation programme (Grant Agreement n$^{\circ}$ 865826), the Deutsche Forschungsgemeinschaft (DFG, German Research Foundation), Grant No. TRR 288-422213477 (Project B03, B06 and B10), Project No. 449386310 and CRC FLAIR. 

\bmhead{Contributions}
H.S. and M.L.T. conceived the project. H.S., F.H., P.B., A.B., M.A.M. and M.L.T. performed the polarized Raman scattering measurements and carried out the Raman data analysis. Q.L., A.H., an
d W.W. conducted the STM experiments. S.K. and C.K. performed the TEM experiments D.F., N.M., and M.M. performed the XRD measurements. A.–A.H. grew the single crystals. H.S. synthesized the intercalated crystals. H.S. and M.L.T. wrote the manuscript with input from all authors.

\bmhead{Competing interests}
The authors declare no competing interests.

\bmhead{Supplementary information}
The online version contains supplementary material available at XXXXXX.

\end{document}